\documentclass[12pt]{article}

\usepackage[dvips]{graphicx}

\def\dspace{\baselineskip = 0.30in}

\def\lapproxeq{\lower .7ex\hbox{$\;\stackrel{\textstyle
<}{\sim}\;$}}
\def\gapproxeq{\lower .7ex\hbox{$\;\stackrel{\textstyle
>}{\sim}\;$}}

\begin{document}

\dspace

\begin{titlepage}
\begin{flushright}
BA-04-05\\
\end{flushright}
\vskip 2cm
\begin{center}
{\Large\bf
$\delta T/T$ and Neutrino Masses in $SU(5)$
}
\vskip 1cm
{\normalsize\bf
Bumseok Kyae\footnote{bkyae@bartol.udel.edu} and
Qaisar Shafi\footnote{shafi@bartol.udel.edu}
}
\vskip 0.5cm
{\it Bartol Research Institute, University of Delaware, \\Newark,
DE~~19716,~~USA\\[0.1truecm]}

%

\end{center}
\vskip 2cm


\begin{abstract}

We implement inflation in a supersymmetric $SU(5)$ model with
$U(1)$ R-symmetry such that the cosmic microwave anisotropy
$\delta T/T$ is proportional to $(M/M_{\rm Planck})^2$,
where $M\sim M_{GUT}=2-3\times 10^{16}$ GeV, the $SU(5)$ breaking scale,
and $M_{\rm Planck}=2.4\times 10^{19}$ GeV.  
The presence of a global $U(1)_X$ symmetry, spontaneously 
broken also at scale $M_{GUT}$, provides an upper bound
$M_{GUT}^2/M_{\rm Planck}\sim 10^{14}$ GeV
on the masses of $SU(5)$ singlet right handed neutrinos,
which explains the mass scale 
associated with atmospheric neutrino oscillations.
The $SU(5)$ monopoles and $U(1)_X$ cosmic strings are inflated away.
Although the doublet-triplet splitting requires fine-tuning, 
the MSSM $\mu$ problem is resolved and dimension five proton decay 
is strongly suppressed. 

\end{abstract}
\end{titlepage}

\newpage



In a class of supersymmetric (SUSY) models of inflation associated 
with some symmetry
breaking $G\rightarrow H$, the cosmic microwave anisotropy $\delta T/T$
turns out to be proportional to $(M/M_{\rm Planck})^2$,
where $M$ denotes the symmetry breaking scale of $G$, and
$M_{\rm Planck}= 1.2\times 10^{19}$ GeV~\cite{hybrid,review}.
Comparison with the determination of $\delta T/T$ by WMAP and
several other experiments~\cite{wmap}
leads to the conclusion~\cite{hybrid,ns}
that $M$ is comparable to $M_{GUT}\sim 2-3\times 10^{16}$ GeV,
the scale at which the three gauge couplings of the minimal supersymmetric
standard model (MSSM) unify.
In such models, the scalar spectral index of density
fluctuations including supergravity (SUGRA) corrections is
$n_s\approx 0.98-1.00$~\cite{ns,sugrainf},
in excellent agreement with observations~\cite{wmap}.
An essential role in the discussion is played
by a global $U(1)_R$ symmetry which helps determine
the form of the tree level superpotential and
has other important phenomenological implications.
For instance, the $Z_2$ subgroup of $U(1)_R$ can be identified
with the ``matter parity'' in the MSSM.
This ensures the absence of rapid proton decay and stability of the LSP.
Examples of $G$ include the groups
$SU(3)_c\times SU(2)_L\times SU(2)_R\times U(1)_{B-L}$
and $SU(4)_c\times SU(2)_L\times SU(2)_R$~\cite{ps}, 
discussed in the context of inflation in Refs.~\cite{LR} and \cite{422}, 
respectively.  

Among supersymmetric grand unified theories (GUTs) 
$SU(5)$~\cite{su5} certainly is the simplest 
with the most direct connection to $M_{GUT}$ given above.
It is therefore natural to try to realize this type of inflationary scenario
in an $SU(5)$ framework~\cite{su5inf}.
The minimal $SU(5)$ model does not incorporate inflation and it also
fails to provide an understanding of neutrino masses inferred
from neutrino oscillations~\cite{nuoscil} and
from cosmological considerations~\cite{wmap}.
Our goal in this paper is to provide a suitable extension
which can overcome both these shortcomings.

We have already indicated the important role to be played
by the $U(1)_R$ symmetry.
Next we consider another symmetry, namely $U(1)_X$,
which also will be important for our analysis.
This is a global symmetry of the minimal model~\cite{u1x} and
its presence prevents the appearance of tree level superheavy masses
($\gapproxeq M_{GUT}$) for the $SU(5)$ singlet
right handed neutrinos.
In our extended model, 
the breaking of $U(1)_X$ triggers the $SU(5)$ breaking at $M_{GUT}$. 
Through non-renormalizable couplings,
the right handed neutrinos acquire masses
of order $M_{GUT}^2/M_P\sim 10^{14}-10^{15}$ GeV, where
$M_X$ and $M_{P}$ ($\equiv M_{\rm Planck}/\sqrt{8\pi}=2.4\times 10^{18}$ GeV)
denote the $U(1)_X$ breaking scale and the reduced Planck mass.
This can yield, via the seesaw mechanism~\cite{seesaw}, 
a light neutrino mass of order
$M_W^2/(10^{14}~{\rm GeV})\sim 10^{-1}$ eV, 
which is the appropriate mass scale for atmospheric neutrino oscillations.
With both 
the global $U(1)_X$ and $SU(5)$ spontaneously broken during inflation,
the $SU(5)$ monopoles and $U(1)_X$ cosmic strings
are inflated away.
At the end of inflation the oscillating fields produce
right handed neutrinos whose out of equilibrium decay lead
via leptogenesis to the observed baryon asymmetry
of the universe~\cite{yanagida}.
The doublet-triplet splitting problem requires, as usual, some fine-tuning. 
However, the mechanism which generates the MSSM $\mu$ term 
after SUSY breaking also strongly suppresses 
dimension five proton decay mediated 
by the superheavy color triplets higgsinos.  
%

%
Let us recall the well known superpotential~\cite{hybrid,lyth},
\begin{eqnarray} \label{simplepot}
W=\kappa S(\phi\bar{\phi}-M^2) ~,
\end{eqnarray}
where $\kappa$ is a dimensionless coupling constant, and
the superfields $\phi$ and $\bar{\phi}$ transform non-trivially under $G$, 
with the gauge invariant combination $\phi\bar{\phi}$ carrying zero $U(1)_R$
charge.
The singlet superfield $S$ and the superpotential $W$ carry
unit $U(1)_R$ charges.
$S$ provides the scalar field that drives inflation.
Note that the $U(1)_R$ symmetry ensures the absence of terms
proportional to $S^2$, $S^3$, etc in the superpotential,
which otherwise could spoil the slow-roll conditions
needed for implementing successful inflation.\footnote{
The issue of SUGRA corrections is more subtle. 
For $W$ given in Eq.~(\ref{simplepot}), and 
the minimal K${\rm\ddot{a}}$hler potential $K$, the flatness condition 
is not spoiled~\cite{review}. However, in some models such as the 
$SU(5)$ case discussed here, additional fields appear 
(from a hidden sector as well as the visible sector), and special 
choices for the K${\rm\ddot{a}}$hler potential may be necessary to control 
SUGRA corrections~\cite{kaehler,newshifted}.
%
}
From $W$, it is straightforward to show that the SUSY minimum
corresponds to non-zero (and equal in magnitude) 
vacuum expectation values (VEVs)
for $\phi$ and $\bar{\phi}$, while $\langle S\rangle =0$,
and therefore $G$ is broken to a subgroup $H$.

An inflationary scenario is realized in the early universe
with $\phi$, $\bar{\phi}$ and $S$ displaced
from their present day minima.
Thus, for $S$ values in excess of the symmetry breaking scale $M$,
the fields $\phi$, $\bar{\phi}$ both vanish,
the gauge symmetry is restored, and a potential energy density proportional
to $\kappa^2M^4$ dominates the universe. With SUSY thus broken, there are
radiative corrections from the $\phi$-$\bar{\phi}$ supermultiplets
that provide logarithmic corrections to the potential
which drives inflation~\cite{hybrid}.
After including the radiative corrections,
the scalar potential from Eq.~(\ref{simplepot}) turns out to give
a scalar spectral index
$n_s\approx 1-1/N$ (for $N\approx 50-60$ e-foldings)~\cite{hybrid}
and the quadrupole anisotropy $\delta T/T$~\cite{422,hybrid}
\begin{eqnarray}\label{T}
\bigg(\frac{\delta T}{T}\bigg)_l\approx \frac{8\pi}{\sqrt{N}}
\bigg(\frac{N_l}{45}\bigg)^{1/2}\bigg(\frac{M}{M_{\rm Planck}}\bigg)^2
x_l^{-1}y_l^{-1}f(x_l^2)^{-1} ~.
\end{eqnarray}
where $x^2_l=(|S|^2/M^2)_l$,
$y_l\approx x_l(1-7/12x_l^2+\cdots)$, $f(x_l^2)^{-1}\approx 1/x_l^2$,
for $S$ sufficiently larger than $M$.
The subscript $l$ is there to emphasize the epoch of horizon crossing.
$N$ indicates the dimensionality of the $\phi$, $\bar{\phi}$ representations,
and $N_l\approx 45-60$ denotes the e-foldings needed to resolve the horizon
and flatness problems.
Comparison of Eq.~(\ref{T}) with the COBE and WMAP data~\cite{wmap} leads
to the conclusion that the gauge symmetry breaking scale $M$ is very close
to $10^{16}$ GeV~\cite{ns}, the SUSY GUT scale inferred from the evolution
of the MSSM gauge couplings~\cite{unif}.
Thus, it is natural to try to realize the above inflationary scenario
within a SUSY GUT framework.
In this paper, we will attempt to provide a realistic scenario
with $G=SU(5)$ and $H=SU(3)_c\times SU(2)_L\times U(1)_Y$.   
%
%

The minimal $SU(5)$ model possesses a global $U(1)_X$ symmetry, where 
$X$ is given by the relation $B-L=X+\frac{4}{5}Y$, which holds for the 
MSSM fields, with $Y$ denoting the hypercharge.  Note that $X$ coincides 
with $B-L$ for the $SU(5)$ singlet right handed neutrinos. 
By imposing a $U(1)_X$ symmetry one prevents the appearance of superheavy
($\gapproxeq M_{GUT}$) masses for the right handed neutrinos 
${\bf 1}_i$ ($i=1,2,3$).
The atmospheric neutrino data and leptogenesis scenario seem to require
right handed neutrino of intermediate masses 
($\lapproxeq 10^{14}-10^{15}$ GeV).
%
%
The masses for ${\bf 1}_i$ can arise from the spontaneous breaking of $U(1)_X$, 
for instance, via the non-renormalizable couplings 
with an extra superfield $\overline{X}$:   
\begin{eqnarray} \label{neutrino}
\frac{y_{ij}}{M_P}~\overline{X}\overline{X}{\bf 1}_i{\bf 1}_j ~, 
\end{eqnarray}
where the $U(1)_X$ charge of $\overline{X}$ is listed in Table I, and 
the dimensionless coupling $y_{ij}$ are 
of order unity or smaller.  
A mass of order $10^{14}-10^{15}$ GeV for the heaviest right handed neutrino 
can nicely explain via the seesaw mechanism 
the atmospheric neutrino mass scale 
($\sqrt{\Delta m^2_{\rm ATM}}\sim 5\times 10^{-2}$ eV). 
This can be achieved with $\langle\overline{X}\rangle\sim M_{GUT}$.  
Hence, it would be desirable to construct a model such that
the $U(1)_X$ and $SU(5)$ breakings are intimately linked.  

If $\overline{X}$ is identified with the inflaton, as will be the case 
in our model, the coupling Eq.~(\ref{neutrino}) can also lead to 
a successful leptogenesis~\cite{yanagida}.       
The inflaton $\overline{X}$ exclusively decays into right handed neutrinos,    
and the reheat temperature $T_r$ turns out to be 
$\sim 10^{-1}M_{\nu^c}$~\cite{422}, 
where $M_{\nu^c}$ is the heaviest right handed neutrino mass allowed by 
the kinematics of the decay process.  
From the gravitino constraint $T_r\lapproxeq 10^9$ GeV~\cite{gravitino}, 
$M_{\nu^c}\sim 10^{10}$ GeV is required.  
Thus, the mass of $\overline{X}$ is constrained as follows:  
\begin{eqnarray} \label{inflatonmass}
m_{\overline{X}}\lapproxeq 10^{14}~{\rm GeV} ~,     
\end{eqnarray}
and the mass(es) of the right handed neutrino(s) lighter 
than $\overline{X}$ should be $\lapproxeq 10^{10}$ GeV.  
In practice, $m_{\overline{X}}$ is taken to be of order $10^{12}-10^{14}$ GeV, 
so that decay into the heaviest right handed neutrino is not allowed.   

The $U(1)_R$ symmetry can forbid
the bare mass term $M{\bf 5}_H{\bf\overline{5}}_H$.
As a consequence, within this sector, the global symmetry
is enhanced to $U(1)_X\times U(1)_{PQ}$, 
with $U(1)_X$ anomaly free.
Motivated by this, we will impose $U(1)_{PQ}$ on the complete model. 
The $U(1)_R$, $U(1)_X$ and $U(1)_{PQ}$ charge assignments for the matter,  
${\bf 5}$-plet Higgs, and the extra superfield are shown in Table I.
\vskip 0.6cm
\begin{center}
\begin{tabular}{|c||ccc|cc|c|} \hline
& ${\bf 10}_i$ & ${\bf\overline{5}}_i$ & ${\bf 1}_i$ &
${\bf 5}_H$ & ${\bf\overline{5}}_H$ & $\overline{X}$
\\
\hline
$U(1)_R$ & $1/2$ & $1/2$ & $1/2$ & $0$ & $0$ & $0$
\\
$U(1)_X$ & $1/5$ & $-3/5$ & $1$ & $-2/5$ & $2/5$ & $-1$
\\
$U(1)_{PQ}$ & $1/5$ & $2/5$ & $0$ & $-2/5$ & $-3/5$ & $0$
\\
\hline
\end{tabular}
\vskip 0.4cm
{\bf Table I~}
\end{center}
Note that with the charge assignments shown in Table I,
the operators leading to
baryon and lepton number violations at low energies
such as ${\bf\overline{5}}_i{\bf 5}_H$,
${\bf 10}_i{\bf\overline{5}}_j{\bf\overline{5}}_k$,
${\bf 10}_i{\bf 10}_j{\bf 10}_k{\bf\overline{5}}_l$,
${\bf 10}_i{\bf 10}_j{\bf 10}_k{\bf\overline{5}}_H$,
${\bf\overline{5}}_i{\bf\overline{5}}_j{\bf 5}_H{\bf 5}_H$,
${\bf\overline{5}}_i{\bf\overline{5}}_H{\bf 5}_H{\bf 5}_H$,
etc. are forbidden in the absence of $U(1)_R$, $U(1)_X$, 
and $U(1)_{PQ}$ breakings.

To reiterate, the $SU(5)$ model we are after should have the following
features.

\noindent $\bullet$ $U(1)_R$ and $U(1)_X$ are suitably utilized
for the desired inflation and neutrino masses.

\noindent $\bullet$ Inflation is associated with the spontaneous breakings of
$SU(5)$ and $U(1)_X$ at $M_{GUT}$.

\noindent $\bullet$ Monopoles and cosmic strings do not pose cosmological
problems.

\noindent $\bullet$ The inflaton satisfying Eq.~(\ref{inflatonmass}) 
decays only into right handed neutrinos via Eq.~(\ref{neutrino}).   

\noindent $\bullet$ The low energy spectrum should coincide
with the MSSM field content.

Consider the following trial superpotentials
\begin{eqnarray} \label{trial2}
&&~~W_{1}=\kappa S\bigg[{\rm Tr}\bigg(\Phi^+\Phi^-\bigg)-M^2\bigg]~,
\\
&&W_{2}=S\bigg[\kappa{\rm Tr}\Phi^2+
\lambda S^+S^--\kappa M^2\bigg]~,
\end{eqnarray}
where $\Phi$'s and $S$'s denote the ${\bf 24}$-plets and singlet superfields, 
respectively.   
We normalize the $SU(5)$ generators $T^i$
such that ${\rm Tr}(T^iT^j)=\delta^{ij}$.
$\kappa$, $\lambda$ are dimensionless couplings, and
the superscripts $\pm$ denote $U(1)_X$ charges of $\pm 1$
for the corresponding Higgs fields.
We assign zero (unit) $U(1)_R$ charges
to $\Phi^{\pm}$, $\Phi$, $S^{\pm}$ ($S$).
From Eq.~(\ref{simplepot}), both $W_1$ and $W_2$ appear to be
viable candidates for inflation 
%
with both $SU(5)$ and $U(1)_X$ broken at the GUT scale.
%
%
However after inflation ends, in either case, unwanted cosmological defects
(monopoles and/or strings) would be produced.
Furthermore, the octet and triplet states 
(and also a pair of ${\bf (3,\overline{2})}_{-5/6}$, 
${\bf (\overline{3},2)}_{5/6}$ states) contained in $\Phi$ ($\Phi^{\pm}$) 
remain `massless' after $SU(5)$ breaking.
These states can acquire masses of order TeV when the $S$ VEV
develops after SUSY breaking.
This would seriously affect the running of the MSSM gauge couplings
and therefore must be avoided.

Before proceeding to the model, let us consider the Higgs superpotential 
$W_3$~\cite{rinvsu5}, which we will encounter 
as a part of the full construction: 
\begin{eqnarray} \label{trial1}
W_{3}=M{\rm Tr}\bigg(\Phi\Phi'\bigg)
+\lambda{\rm Tr}\bigg(\Phi\Phi'\Phi'\bigg)~,
\end{eqnarray}
where $\Phi$ ($\Phi'$) is a ${\bf 24}$-plet superfield with 
unit (zero) $U(1)_R$ charge. 
Compared to minimal $SU(5)$, an additional ${\bf 24}$-plet Higgs 
$\Phi$ is introduced to realize the $U(1)_R$ symmetry.   
While $\langle\Phi\rangle=0$ in the SUSY minimum, 
the scalar component of $\Phi'$ acquires a VEV ($\sim M/\lambda$)
along the $SU(3)_c\times SU(2)_L\times U(1)_Y$ singlet direction.
The ${\bf (8,1)}_0$, ${\bf (1,3)}_0$, and ${\bf (1,1)}_0$ components in $\Phi$ 
and $\Phi'$ pair up and become superheavy. 
While ${\rm Im}[{\bf (3,\overline{2})}_{-5/6}]$ and
${\rm Im}[{\bf (\overline{3},2)}_{5/6}]$ obtain superheavy masses from
the D-term potential, 
the goldstone modes ${\rm Re}[{\bf (3,\overline{2})}_{-5/6}]$,   
${\rm Re}[{\bf (\overline{3},2)}_{5/6}]$ contained in $\Phi'$ 
are absorbed by the massive gauge bosons when $SU(5)$ is broken.   
Hence, the components ${\bf (3,\overline{2})}_{-5/6}$ and  
${\bf (\overline{3},2)}_{5/6}$ from $\Phi$ remain `massless.' 
To make them superheavy,  
a coupling such as $\langle\Sigma_{-1}\rangle{\rm Tr}(\Phi\Phi)$  
with a large VEV for $\Sigma_{-1}$ is needed, 
where $\Sigma_{-1}$ denotes a $SU(5)$ singlet field  
carrying a $U(1)_R$ charge of $-1$.  
%
%
%

For a realization of inflation and $U(1)_X$ in $SU(5)$, 
we introduce some $SU(5)$ singlets and adjoint Higgs superfields 
with $U(1)_R$, $U(1)_X$ and $U(1)_{PQ}$ charges shown in Table II.
\vskip 0.6cm
\begin{center}
\begin{tabular}{|c||ccc|ccc|cc|} \hline
 & $S$ & $S^+$ & $S^-$ & $Z$ & $P$ & $Q$  
& $\Phi$ & $\Phi^+$ 
%
\\
\hline
$SU(5)$ & {\bf 1} & {\bf 1} & {\bf 1} & {\bf 1} & {\bf 1} & {\bf 1} 
& {\bf 24} & {\bf 24} 
\\
$U(1)_R$ & $1$ & $0$ & $0$ & $2$ & $-1$ & $-1$ & $1$ & $0$ 
\\
$U(1)_X$ & $0$ & $1$ & $-1$ & $-3$ & $2$ & $4$ & $-2$ & $1$ 
\\
$U(1)_{PQ}$ & $0$ & $0$ & $0$ & $0$ & $0$ & $0$ & $0$ & $0$ 
\\
\hline
\end{tabular}
\vskip 0.4cm
{\bf Table II~}
\end{center}
We identify $S^-$ with $\overline{X}$ in Table I.     
To make $U(1)_X$ anomaly free,  
one could introduce, for instance,  
an additional singlet ($T$) and an adjoint Higgs ($\Phi^+_{1/2}$) 
with $U(1)_X$ charges $-3$ and $+1$, respectively.
We assign a $U(1)_R$ charge of $1/2$ to $\Phi^+_{1/2}$, 
and a proper non-zero $U(1)_{PQ}$ charge to $T$.   
Their presence does not affect the inflationary scenario 
in any significant manner.      
The relevant renormalizable superpotential is given by
\begin{eqnarray} \label{superpot}
&&W_{\rm infl}=\kappa_1 S\bigg[S^+S^--M^2\bigg]+ 
Z\bigg[\kappa_2PS^+-\kappa_3QS^-\bigg] 
\\
%
%
&&~+\lambda Q{\rm Tr}\bigg(\Phi\Phi\bigg)
+\alpha S^+{\rm Tr}\bigg(\Phi\Phi^+\bigg)
+\beta {\rm Tr}\bigg(\Phi\Phi^+\Phi^+\bigg) ~,       
\nonumber 
\end{eqnarray}
%
where $\alpha$, $\beta$, $\lambda$, $\kappa$'s are dimensionless coefficients.
Note that through a suitable redefinition of the superfields,
the parameters in Eq.~(\ref{superpot}) can all be made real. 
Several comments are in order here.  
The $\kappa_1$ terms in Eq.~(\ref{superpot}) drives inflation 
accompanied by the spontaneous breaking of $U(1)_X$.  
Following Ref.~(\cite{ns}), we assume that $\kappa_1$ is of order 
$10^{-3}-10^{-2}$, which keeps the SUGRA corrections under control.  
The last two terms in Eq.~(\ref{superpot}) resemble $W_3$ discussed above, 
and therefore the $SU(5)$ breaking is triggered by a superlarge VEV of $S^+$.   
From the $\lambda$ term,  the ${\bf (3,\overline{2})}_{-5/6}$ and 
${\bf (\overline{3},2)}_{5/6}$ components 
contained in $\Phi$ obtain superheavy masses with $Q$ acquiring a superlarge 
VEV via the $\kappa_2$, $\kappa_3$ terms in Eq.~(\ref{superpot}).    
%
%
%
[Through a non-renormalizable term $S^-S^-\Phi^+_{1/2}\Phi^+_{1/2}/M_P$, 
the component fields contained in $\Phi^+_{1/2}$ 
all acquire the same heavy mass. 
Thus, $\Phi^+_{1/2}$ leaves intact 
the unification of the MSSM gauge couplings.   
The VEV of $\Phi^+_{1/2}$ turns out to vanish at the SUSY minimum.]      
Consequently, without leaving any unwanted light fields,  
$SU(5)\times U(1)_X$ is broken to the MSSM gauge symmetry.  

From Eq.~(\ref{superpot}), one derives the F-term scalar potential:
\begin{eqnarray} \label{pot}
&&~~V_{\rm infl}=\bigg|\kappa_1 S^+S^--\kappa_1 M^2\bigg|^2
+\bigg|\kappa_2PS^+-\kappa_3QS^-\bigg|^2
+\bigg|\kappa_1SS^+-\kappa_3ZQ\bigg|^2
\nonumber \\
&&~
+\bigg|\kappa_1SS^-+\kappa_2ZP
+\alpha\sum_i\Phi_{i}\Phi^+_i\bigg|^2
+\bigg|\kappa_2ZS^+\bigg|^2
+\bigg|\kappa_3ZS^-
-\lambda\sum_i\Phi_{i}\Phi_{i}\bigg|^2
\\
&&+\sum_i\bigg(\bigg|2\lambda Q\Phi_{i}
+\alpha S^+\Phi^+_i+\beta \sum_{j,k}d^{ijk}\Phi^+_j\Phi^+_k\bigg|^2
+\bigg|\alpha S^+\Phi_{i}
+2\beta \sum_{j,k}d^{ijk}\Phi_{j}\Phi^+_k\bigg|^2
\bigg)
~,~~
\nonumber
\end{eqnarray}
%
%
%
where $d^{ijk}\equiv {\rm Tr}(T^iT^jT^k)$, and $\Phi_{i}^{(+)}$ 
($i=1,2,3,\cdots, 24$) denotes the component fields 
of the ${\bf 24}$-plet Higgs. 
%
At the SUSY minimum of Eq.~(\ref{pot}), the VEVs of $S$, 
$Z$, and $\Phi$ vanish
\begin{eqnarray}
\langle S\rangle=\langle Z\rangle=\langle\Phi_{i}\rangle
=0 ~, ~~~~~i=1,2,3,\cdots, 24 ~.
\end{eqnarray}
In the presence of soft SUSY breaking terms these fields can acquire  
VEVs of as much as a TeV.   
On the other hand, $S^{\pm}$, $P$, $Q$, $\Phi^+$ can develop VEVs
such that $SU(5)\times U(1)_X$ is spontaneously broken
to the MSSM gauge group,
\begin{eqnarray} \label{vev}
&&\langle S^+\rangle\langle S^-\rangle=M^2 ~,~~~
\kappa_2\langle P\rangle\langle S^+\rangle
=\kappa_3\langle Q\rangle\langle S^-\rangle ~, 
\\
&&~~~~~~~~|\langle\Phi^+_8\rangle|
=\bigg|\frac{\alpha\langle S^+\rangle}{\beta d^{888}}\bigg|
\equiv M_{GUT}\sim M~, 
\label{phi}
\end{eqnarray}
where $\Phi^{+}_8$ is the MSSM singlet contained in $\Phi^+$ 
($\langle\Phi^+_i\rangle=0$ for $i\neq 8$), 
and $d^{888}$ is given by $\frac{1}{\sqrt{30}}$.
%
%
By performing an appropriate $U(1)_X$ transformation,
we can rotate away the phase of $\langle S^+\rangle$. 
We identify $|\langle\Phi_8^+\rangle|$ 
with $M_{GUT}\approx 2-3\times 10^{16}$ GeV. 
We assume $\langle S^+\rangle\sim\langle S^-\rangle\sim
M\sim M_{GUT}$, and $\langle P\rangle\sim\langle Q\rangle\sim 10^{17}$ GeV   
with $U(1)_R$ broken to $Z_2$ (``$R$-parity'').  
The VEVs of $P$ and $Q$ could be stabilized 
by including the soft terms and the SUGRA corrections 
to the scalar potential.   We will see later that
$\kappa_{2,3}$ should be of order $10^{-4}-10^{-3}$,
while $\lambda$, $\alpha$, $\beta$ are of order unity.
%
%
%
%

With $U(1)_X$ broken at the $M_{GUT}$ scale,
we get an upper bound on the right handed neutrino Majorana mass 
by identifying $S^{-}$ with $\overline{X}$ in Eq.~(\ref{neutrino}):
\begin{eqnarray}
\frac{\langle S^-S^-\rangle}{M_{P}}
\sim 10^{14}-10^{15}~{\rm GeV}.
\end{eqnarray}
Assuming the heaviest right handed neutrino of this mass,  
and with a Dirac neutrino mass 
for the third family of order the electroweak scale,
we find a light neutrino mass 
$M_W^2/(10^{14}~{\rm GeV})\sim\sqrt{\Delta m^2_{\rm ATM}}
\sim 5\times 10^{-2}$ eV.  

Next let us discuss how to implement doublet-triplet splitting.
With some additional superfields,
whose quantum numbers appear in Table III,
\vskip 0.6cm
\begin{center}
\begin{tabular}{|c||cc||cccc||cc|} \hline
 & $H$ & $\overline{H}$ 
& $\Delta$ & $C$ & $A$ & $B$
& $N$ & $\overline{N}$
%
\\
\hline
$SU(5)$ & ${\bf 5}$ & ${\bf \overline{5}}$ 
& ~~${\bf 1}$~ & ${\bf 1}$ & ${\bf 1}$ & ${\bf 1}$ 
& {\bf 1} & {\bf 1}
%
\\
$U(1)_R$ &
$1$ & $1$ 
& $2$ & $-1$ & $-1$ & $0$ 
& ~$1/2$~ & $0$
%
\\
$U(1)_X$ &
$-7/5$ & $-3/5$ 
& $0$ & $0$ & $2$ & $-2$
& $0$ & $0$ 
\\
$U(1)_{PQ}$ &
$3/5$ & $2/5$ 
& $3/2$ & $-3/2$ & $-1$ & $-1/2$
& $1/2$ & $-1/2$ 
%
\\
\hline
\end{tabular}
\vskip 0.4cm
{\bf Table III~}
\end{center}
the relevant superpotential is
\begin{eqnarray} \label{dt}
W_H&=&y_aH\bigg[S^+ +a\Phi^+\bigg]
{\bf\overline{5}}_H
+y_b\overline{H}\bigg[S^+ +b\Phi^+\bigg]{\bf 5}_H
\nonumber \\
&&+\Delta\bigg[y_cMC-y_dAB\bigg]+y_eAH\overline{H}
\\
&&+\frac{y_\mu}{M_P}NN{\bf 5}_H{\bf\overline{5}}_H
+\frac{y_n}{M_P}NN\overline{N}\overline{N}~, 
\nonumber 
\end{eqnarray}
whose presence does not affect the conclusions based on Eq.~(\ref{superpot}).
[The $y$'s  and $a$, $b$ denote dimensionless couplings.]   
From the $y_c$, $y_d$ terms and the soft terms, 
$C$, $A$, $B$ can develop VEVs of order $M$, 
while $\langle\Delta\rangle$ is of order the gravitino mass scale $m_{3/2}$. 
Thus, $U(1)_{PQ}$ is also broken at $M_{GUT}$, but presumably this is 
acceptable within an inflationary cosmology~\cite{linde}.  
The $y_e$ term in Eq.~(\ref{dt})
ensures that the additional ${\bf 5}$-plets acquire superheavy masses.
The VEVs $\langle S^+\rangle$, $\langle\Phi_8^+\rangle$
would make ${\bf \overline{5}}_H$, ${\bf 5}_H$ superheavy.
Thus, two fine-tunings, 
$a=b=\sqrt{10/3}\langle S^-\rangle/\langle\Phi_8^-\rangle$  
are necessary to obtain a pair of light Higgs doublets
from ${\bf\overline{5}}_H$, ${\bf 5}_H$.
The non-renormalization theorem in SUSY ensures that
such fine-tunings are stable against radiative corrections.
We assume that $y$'s and $a$ ($=b$) are of order unity.

Due to the presence of SUSY breaking ``$A$-terms,''
the scalar components of $N$, $\bar{N}$ acquire intermediate scale VEVs,
$\sim\sqrt{m_{3/2}M_{P}}\sim 10^{10}$ GeV. 
Consequently, 
%
a $\mu$ parameter ($\equiv y_\mu\langle NN\rangle/M_P$) 
of order $m_{3/2}$ is naturally generated~\cite{mupara}.
The presence of $U(1)_{PQ}$ resolves the strong CP problem~\cite{axion}.

The non-zero VEVs of $N$, $\overline{N}$ also break    
the $Z_2$ symmetry ($\subset U(1)_R$), 
which can lead to a  cosmological domain wall problem.     
We therefore assume that the VEVs  
$\langle N\rangle$, $\langle\overline{N}\rangle$ 
develop before or during inflation, so that 
the domain walls are inflated away.    
With ${\bf 10}_i{\bf\overline{5}}_j{\bf\overline{5}}_H$, 
${\bf\overline{5}}_i{\bf 5}_H\langle N\overline{N}S^+\rangle/M_P^2$, 
and the induced $\mu$ term in Eq.~(\ref{dt}),  
the trilinear operator ${\bf 10}_i{\bf\overline{5}}_j{\bf\overline{5}}_k$ 
leading to ``$R$-parity'' violation in the MSSM is generated  
with a suppression factor $\mu^2/(M_PM_{GUT})$.
If one requires an absolutely stable LSP,  
one could introduce a $Z_2$ ``matter parity.'' 

Before discussing some aspects of inflation, let us point out 
an important consequence of the model related to proton decay. 
Because of the absence of a direct coupling 
${\bf 5}_H{\bf\overline{5}}_H$ in the $U(1)_R$ symmetric limit,  
the higgsino mediated dimension five operator relevant for proton decay,  
${\bf 10}_i{\bf 10}_j{\bf 10}_k{\bf\overline{5}}_l$, is suppressed by 
$\mu/M_{GUT}^2$ ($\sim M_W/M_{GUT}^2$), which makes it harmless.
The higher dimensional operator 
${\bf 10}_i{\bf 10}_j{\bf 10}_k{\bf\overline{5}}_l
\times\langle S^-S^-A\rangle/M_{P}^4$ 
$\sim {\bf 10}_i{\bf 10}_j{\bf 10}_k{\bf\overline{5}}_l\times 10^{-7}/M_P$  
is also suppressed.    
Thus, proton decay is expected to proceed via the superheavy gauge bosons, 
with life time $\sim 10^{34}-10^{36}$ yrs.   

In this model, inflation is implemented by assuming that in the early universe 
$\langle S\rangle\gapproxeq M$ with $\langle P\rangle\sim 10^{17}$ GeV.   
As a result, the VEVs of $S^{+}$, $Q$, and $\Phi^+$ vanish,
and a vacuum energy density proportional to $\kappa_1^2 M^4$ dominates
the universe, as in inflation based on Eq.~(\ref{simplepot}).   
Inflation is driven by the radiatively generated logarithmic 
inflaton potential.   
With $\langle\Phi\rangle$  
as well as $\langle S^-\rangle$, $\langle Z\rangle$ non-zero 
during inflation~\cite{newshifted,5d422}, $SU(5)\times U(1)_X$ is broken
to $SU(3)\times SU(2)\times U(1)$,   
and the $SU(5)$ monopoles and $U(1)_X$ strings are inflated away.
For simplicity, we assume that
$\kappa_2\langle P\rangle\sim\frac{1}{10}\kappa_1M
>>\langle \Phi\rangle$, $\langle S^-\rangle$, $\langle Z\rangle$ so that 
Eq.~(\ref{T}) with $N=2$ still approximately holds.   
%
%
It turns out that with inclusion of soft SUSY breaking terms, $S^+$, $\Phi^+$ 
acquire VEVs during inflation that are proportional to $m_{3/2}$, the SUSY 
breaking scale. As inflation ends, $S^+$ and $\Phi^+$ develop GUT scale VEVs, 
while $\Phi$ is driven to zero, so that $SU(5)$ is broken 
to the MSSM gauge group.  

With inflation over, the inflaton fields oscillate about their SUSY minima.
With the masses of $S^+$ ($=|\kappa_1\langle S^-\rangle |
\sim|\kappa_2\langle P\rangle |\lapproxeq 10^{14}$ GeV) and 
$\Phi_8^+$ ($=|\alpha\langle S^+\rangle|\lapproxeq M_{GUT}$) 
smaller than the masses of the triplet Higgs contained in 
${\bf 5}_H$, ${\bf\overline{5}}_H$ 
($\langle S^-\rangle\sim\langle \Phi_8^+\rangle=M_{GUT}$), 
$S^+$, $\Phi^+$ can not decay through the terms in Eq.~(\ref{dt})
into the triplet Higgs.
A linear combination $S^+-c\sqrt{\frac{3}{10}}\Phi_8^+$ $(\equiv\Psi)$,
where $c\equiv a=b=\sqrt{10/3}~
[\langle S^+\rangle/\langle\Phi_8^+\rangle]_{\rm min}$,
couples to the doublets in ${\bf 5}_H$, ${\bf\overline{5}}_H$.
We note that the VEV of $\Psi$ vanishes during and after inflation.
Indeed, as $S$ (and $Z$) rolls down to the origin,
$S^+$ and $\Phi_8^+$ also roll down, from the origin
to their present values. 
If $\Psi$ remains zero throughout inflation,\footnote{
We assume that this can be realized through a judicious choice of
parameters in the potential in Eq.~(\ref{pot}). 
This would be impossible had we assigned a zero $U(1)_X$ charge to $\Phi^+$,  
and the ${\bf 5}$-plet Higgs masses were given in terms of  
a mass parameter $M$ and $\langle\Phi^+\rangle$.  
}
the field $S^-$ exclusively decays into
right handed neutrinos and sneutrinos
via $S^-S^-{\bf 1}_i{\bf 1}_j/M_{P}$ 
and the superpotential couplings in Eq.~(\ref{superpot}).
With $\kappa_{1}\lapproxeq 10^{-2}$, $\kappa_3\lapproxeq 10^{-3}$, 
$\langle S^+\rangle\sim 10^{16}$ GeV, and $\langle Q\rangle\sim 10^{17}$ GeV, 
the inflaton $S^-$ fulfills Eq.~(\ref{inflatonmass}).   
The subsequent out of equilibrium decay of the right-handed neutrinos
yields the observed baryon asymmetry
via leptogenesis~\cite{yanagida}. 

Before concluding, one may inquire about implementing inflation
in five dimensional (5D) $SU(5)$ models
which have attracted much recent attention
because of the relative ease with which the doublet-triplet problem
is resolved and higgsino mediated dimension five proton decay
is eliminated~\cite{5d}.
Following Refs.~\cite{5dinf,5d422}, inflation with
$\delta T/T\propto (M_{X}/M_{\rm Planck})^2$ can be realized in this case,
where $M_{X}$ ($\sim 10^{16}$ GeV) denotes the $U(1)_X$ breaking scale.  
Thus, the desired atmospheric neutrino mass can be realized also
in 5D $SU(5)$.

In conclusion, we have shown how inflation can be realized in SUSY
$SU(5)$ with $U(1)_R$ symmetry playing an essential role.
We have also discussed how neutrino masses can be understood in this setting
by exploiting a global $U(1)_X$ symmetry also broken 
at a scale close to $M_{GUT}$.
This inflationary model also possesses some interesting phenomenology. 
In particular, the troublesome dimension five proton decay mediated by 
higgsino exchange in minimal $SU(5)$ is strongly suppressed.    
It would be of some interest to extend the discussion to larger GUTs 
such as $SO(10)$ and $E_6$.  
%
%

\vskip 0.3cm
\noindent {\bf Acknowledgments}

\noindent
The work is partially supported
by DOE under contract number DE-FG02-91ER40626.

\end{document}